\documentclass[10pt,conference]{IEEEtran} \newcommand{\figsza}{0.9} \newcommand{\figszb}{0.95}

\usepackage[cmex10]{amsmath}
\usepackage{amsfonts}
\usepackage{amssymb}
\usepackage{amsopn}
\usepackage{graphicx}
\usepackage{subfigure}
\usepackage{epsfig}
\usepackage{color}
\usepackage{setspace}

\graphicspath{{fig/}}

\newcommand{\ud}{\,\mathrm{d}}

\newcommand{\tr}{\text{Tr }}
\newcommand{\bo}[1]{\boldsymbol{#1} }
\newcommand{\sups}[1]{\ensuremath{^{\textrm{#1}}}}

\makeatletter
\def\blfootnote{\xdef\@thefnmark{}\@footnotetext}
\makeatother

\begin{document}

%
\title{Spatial Whitening Framework for Distributed Estimation}

\author{\IEEEauthorblockN{Swarnendu Kar and Pramod K. Varshney}
\IEEEauthorblockA{Dept. of Electrical Engineering and Computer Science\\
Syracuse University\\
Syracuse, NY,  13244, USA\\
Email: \{swkar,varshney\}@syr.edu}
\and
\IEEEauthorblockN{Hao Chen}
\IEEEauthorblockA{College of Engineering\\
Boise State University\\
Boise, ID 83725\\
Email: haochen@boisestate.edu}
}

\maketitle

\begin{abstract}
Designing resource allocation strategies for power constrained sensor network in the presence of correlated data often gives rise to intractable problem formulations. In such situations, applying well-known strategies derived from conditional-independence assumption may turn out to be fairly suboptimal. In this paper, we address this issue by proposing an adjacency-based spatial whitening scheme, where each sensor exchanges its observation with their neighbors prior to encoding their own private information and transmitting it to the fusion center. We comment on the computational limitations for obtaining the optimal whitening transformation, and propose an iterative optimization scheme to achieve the same for large networks. We demonstrate the efficacy of the whitening framework by considering the example of bit-allocation for distributed estimation.
\end{abstract}

\section{Introduction}
Wireless sensor networks consist of spatially distributed noisy sensors that cooperatively monitor environmental conditions. Since the individual sensor nodes are characterized by limited energy, bandwidth and computational capability, the task of the fusion center (FC) is to make accurate inference about the phenomenon by requesting as little information from the sensor nodes as possible \cite{Rib06}. Depending on the particular application and set of constraints, the FC often has to adopt smart strategies to collect and process data \cite{Varshney97}. While the design of optimum strategies in some cases is relatively easy under the assumption of \emph{conditional independence}\footnote{Here, `independence' refers to the statistical independence of sensor data \emph{conditioned} on the parameter of interest. For additive Gaussian observation noise, this is equivalent to the covariance matrix of noise being diagonal. The observations are still marginally dependent, since they are observing the same parameter.} across sensors, it is well known that the design gets harder and sometimes the optimum strategy is intractable when correlation has to be taken into account \cite{Krasnopeev05}. In particular, when the sensors are geographically close, they are expected to possess significant correlation among themselves and the optimum strategies derived for the independent case will no longer be optimal. In this paper, we introduce a framework called \emph{spatial whitening} (to be formalized later) to deal with this problem. \blfootnote{This research was partially supported by the National Science Foundation under Grant No. 0925854 and the Air Force Office of Scientific Research under Grant No. FA-9550-10-C-0179.} \blfootnote{This is the postprint of a paper presented at the 2011 4th IEEE International Workshop on Computational Advances in Multi-Sensor Adaptive Processing (CAMSAP), 13-16 Dec. 2011, San Juan, Puerto Rico, USA.} \blfootnote{http://dx.doi.org/10.1109/CAMSAP.2011.6136007} 

Our framework stems from this idea: If two sensors in a network are highly correlated, they are also likely to be spatially close, which means that they should be able to communicate and exchange information among themselves in a relatively inexpensive manner (avoiding routing overheads and long distance communications). Each sensor in the network can now use the information from neighboring nodes to achieve a \emph{local whitening transformation}. If each of such local transformations can be coordinated, one can aim to achieve \emph{global whitening}, and the transformed observations can then be transmitted to the FC using optimum encoding strategies (for inference, resource allocation, etc) that were derived for conditionally independent scenarios. Hence, this two-stage (whitening followed by encoding) framework potentially enables the use of several earlier known results in the presence of correlated noise.

We introduce the log-determinant divergence based formulation of spatial whitening in Section \ref{prob:statement}. To illustrate the potential usage of this framework, we employ the problem of distributed parameter estimation \cite{Rib06}, where several sensor nodes quantize their individual observations before sending them to FC. The goal is to minimize the expected distortion of the estimated parameter subject to a constraint on the total number of bits transmitted to the FC. We demonstrate that an optimal strategy for bit allocation (derived for independent scenario \cite{Krasnopeev05}) delivers increasingly better performance with increasing degree of whitening.

The whitening transformation described in this paper requires local message passing which is certainly not without cost. However, in this paper, we assign no cost to whitening, acknowledging fully that any actual implementation of a system would have to consider the tradeoff between the benefits of whitening and the cost of it. Investigations on this tradeoff is a worthy topic for future research.

The concept of whitening, in general, has mostly been addressed in a global framework till now. It is well known that the Karhunen-Lo\`eve Transform (KLT) \cite{Gastpar06} (also referred to as Principal Components Analysis, PCA) of a random vector with covariance matrix $\bo \Sigma=\bo U\bo \Lambda \bo U^T$ provides the unique whitening transformation ($\bo U^T$) that is also orthogonal. However, PCA is ill suited for our problem, since those whitening transformations are not local, while the orthogonality property serves no additional purpose. The Cholesky decomposition $\bo \Sigma=\bo L\bo L^T$, which provides the unique lower triangular whitening transformation ($\bo L^{-1})$, also requires non-local transformations. Moreover, the lower-triangular property imposes a tree-type dependence structure while in fact there is no natural ordering of spatially correlated data \cite{Zhu09}. Other sparsity-inducing decompositions like Sparse-PCA \cite{Tibshirani06} and vector Sparse-PCA \cite{Ulfarsson11} are \emph{exploratory}\footnote{The placeholders for non-zero coefficients are not known/specified beforehand.} in nature, which means that the resulting transformations are not guaranteed to be local. In \cite{Venkatesan04}, a hardware-friendly technique was proposed to achieve generic spatial whitening transformations that were also global in scope. In distributed-KLT \cite{Gastpar06}, individual nodes observe non-overlapping portions of a random vector and perform dimensionality-reduction (without collaboration with neighbors) for optimum reconstruction at the FC. 

Local communication among sensors has mostly been used to address \emph{in-network} inference problems till now. Distributed consensus problems \cite{Dimakis10} aim at designing iterative message passing schemes in order to compute (some linearly weighted) average at all the nodes. Graphical model based problems involve the selection of structured inverse-covariance matrices (example 2.5 in \cite{Vandenberghe98}) and the subsequent design of message passing schemes for (posterior) belief computation \cite{Cetin06} at all nodes. In this paper, we address the fixed FC based problem - where the inference is performed at the FC rather than inside the network.

Our primary contribution in this paper is the formulation of a whitening framework that harnesses (some minimal) local communication among sensors for efficient resource allocation in fixed FC applications.


\section{Problem Statement} \label{prob:statement}
We consider $N$ sensors in a network that is observing an unknown, deterministic, scalar parameter of interest $\theta$ in the presence of zero-mean, correlated Gaussian noise with covariance $\bo \Sigma$. Hence the sensor observations $\bo x=[x_1,x_2,\ldots,x_N]$ follow
\begin{align}
\bo x \sim \mathcal N(\bo 1 \theta,\bo \Sigma), \quad \bo 1\triangleq [1\cdots 1]^T\in \mathbb R^N.
\end{align}
Note that the sensor observations $x_k$-s are conditionally-independent when $\bo \Sigma$ is a diagonal matrix. Let the neighborhood structure among the various nodes be represented by the $N\times N$ adjacency matrix $\bo A, A_{ij}\in \{0,1 \}$, which is expected to be sparsely populated. Entries $A_{ij}=1$ signify that node $i$ is a neighbor of node $j$. A low-cost link for local communication is assumed to be available between two neighboring links. Since each node is trivially connected to itself, $A_{ii}=1$. We denote the set of all $\bo A$-sparse matrices as
\begin{align}
\mathcal S_A\triangleq \{\bo W\in\mathbb R^{N\times N}:W_{ij}=0 \text{ if } A_{ij}=0 \}.
\end{align}
Note that because $\bo A$ is the adjacency matrix, all linear transformations of the form
\begin{align}
\widetilde{\bo x}=\bo W \bo x\sim\mathcal N(\bo W \bo 1 \theta,\bo W \bo \Sigma \bo W^T),\quad \bo W\in \mathcal S_A,
\end{align}
can be realized relatively inexpensively through local data transmissions, i.e., node $k$ realizes the transformation $\widetilde x_k=\sum_{j\in \mathcal{N}_k} W_{kj} x_j$ by collaborating with its set of neighbors $\mathcal N_k\triangleq \{j_1,j_2,\ldots , j_{|\mathcal N_k|}\}$, i.e., all the columns-indices of $\bo A$ such that $A_{k,j_i}=1$.

The goal is to find the optimal mean-preserving, whitening transformation, i.e., one for which $\bo W \bo 1=\bo 1$, and $\bo W \bo \Sigma \bo W^T$ is as \emph{near} to some diagonal matrix as possible. The mean-preserving condition ensures that the problem framework is preserved, i.e., any resource allocation algorithm previously designed for the observation domain $\bo x$ is applicable to new transformed domain $\widetilde{\bo x}$. The whitening condition helps induce conditional independence across sensors (in some optimal sense). We chose the log-determinant divergence \cite{Dhillon07} as our metric for matrix-\emph{nearness}, a point that we will elaborate later. The idea is that the nodes can use (optimally) whitened observations $\widetilde x_k$ (instead of original correlated observations $x_k$) as the information to be encoded and relayed to the FC. This way an encoding strategy that was derived using conditional independence assumption across sensors can be used to enhance the performance of the system. We will consider the application of optimal encoding for distributed estimation in Section \ref{sec:bit} and show the resulting improvement in performance due to the two-stage processing. But before that we describe our approach towards finding the optimum whitening transformation and comment on the computational aspects.

In the domain of symmetric positive-definite $N\times N$ matrices, the log-determinant divergence of $\bo P$ from $\bo Q$ is defined \cite{Dhillon07} as
\begin{align}
\mathcal L(\bo P;\bo Q)\triangleq \tr \bo Q^{-1}\bo P-\log \det \bo P-N+\log \det \bo Q. \label{def:logdet:div}
\end{align}
It is well known that $\mathcal L(\bo P;\bo Q)$ is a Bregman-divergence \cite{Dhillon07} and hence convex in $\bo P$ for any fixed $\bo Q$. Also $\mathcal L(\bo P;\bo Q)\ge 0$ for all $\bo P$ and $\bo Q$ with equality if and only if $\bo P=\bo Q$. We formulate the spatial whitening problem as finding an $\bo A$-sparse, mean-preserving transformation $\bo W$ and a diagonal matrix (with positive entries) $\bo D$ such that the divergence $\mathcal L(\bo W \bo \Sigma \bo W^T;\bo D)$ is minimized,
\begin{align}
\min_{\bo W, \bo D} \quad \mathcal L(\bo W \bo \Sigma \bo W^T;\bo D) \quad \text{s.t.} \quad & \bo W\in \mathcal S_A, \bo W\bo 1=\bo 1. \label{maxl1}
\end{align}
We note from definition \eqref{def:logdet:div} that
\begin{align}
\mathcal L(\bo W \bo \Sigma \bo W^T;\bo D)=\mathcal L(\bo D^{-\frac{1}{2}} \bo W \bo \Sigma \bo W^T\bo D^{-\frac{1}{2}};\bo I), \label{equiv:div}
\end{align}
where $\bo I$ is the identity matrix. Using \eqref{equiv:div}, we obtain an equivalent formulation of \eqref{maxl1},
\begin{align}
&\min_{\bo Z} \quad \mathcal L(\bo Z \bo \Sigma \bo Z^T;\bo I) \quad \text{s.t.} \quad \bo Z\in \mathcal S_A,
 \label{maxl} \\
&\text{where} \quad \bo W=\delta^{-1}(\bo Z\bo 1) \bo Z, \quad \bo D=\delta^{-2}(\bo Z\bo 1), \label{def:WD:Z}
\end{align}
where $\delta(\cdot)$ is the diagonalization\footnote{Function $\bo X=\delta(\bo x)$ is defined as $\delta:\mathbb R^N\rightarrow \mathbb R^{N\times N}$ such that $\bo x$ corresponds to the diagonal elements of $\bo X$, other elements being zero.} operator. We note that \eqref{maxl} is a significantly simplified re-formulation of \eqref{maxl1}. Using \eqref{def:logdet:div}, we define the cost function w.r.t. $\bo Z$ as
\begin{align}
l(\bo Z)&\triangleq \mathcal L(\bo Z \bo \Sigma \bo Z^T;\bo I) =\tr \bo Z \bo \Sigma \bo Z^T-\log \det \bo Z \bo Z^T+c_0, \label{def:cost:fn}
\end{align}
where $c_0\triangleq-N-\log \det \bo \Sigma$ is a constant. We refer to \eqref{maxl} as the \emph{log-determinant divergence based spatial whitening} problem. If the cardinality of non-zero elements of $\bo A$ is $\text{nz}(\bo A)\le N^2$, then \eqref{maxl} is an optimization problem  in $\mathbb R^{\text{nz}(\bo A)}$.

Since $\bo Z$ is not restricted to the set of symmetric positive-definite matrices (denoted by $\bo S^{++}$), our objective function \eqref{def:cost:fn} does not inherit the convexity property of well known \emph{max-det} problems \cite{Vandenberghe98}. Neither does the first-order gradient condition, written in matrix-derivative notations \cite{Magnus99},
\begin{align}
\frac{\ud l(\bo Z)}{\ud \bo Z}=2 (\bo Z\bo \Sigma -\bo Z^{-T})\circ A=0, \label{stat:cond}
\end{align}
where $\circ$ denotes the element-wise (or Hadamard) product, lend itself to any known closed-form solution except in the trivial situation when $\bo A$ is the all-$1$ matrix (in which case, $\bo Z \bo \Sigma \bo Z^T=\bo I$, and any orthogonal multiple of the Cholesky factor $\bo L^{-1}$ is a solution for $\bo Z$).
In the next section, we provide an iterative algorithm that finds (locally) optimal solutions to problem \eqref{maxl}. Multiple runs using \emph{good}\footnote{$l(\bo Z)$ is convex in the smaller subset $\mathcal S_A\bigcap\bo S^{++}$, the minima within which can be efficiently computed and considered a \emph{good} starting point.} starting points must be used to mitigate the local-maxima problem and obtain a satisfactory solution. It may be noted here that in most of existing literature, matrix factorization problems of this nature (involving sparsity/structure) are inherently non-convex and can only guarantee locally optimal solutions  \cite{Gastpar06}, \cite{Tibshirani06}, \cite{Ulfarsson11}.

\section{Iterative Algorithm for Spatial Whitening} \label{sec:pbpo}
In our iterative approach to solving problem \eqref{maxl}, we update each row of elements in $\bo Z$ to achieve the optimum decrement in divergence, while keeping the rest of the matrix unchanged. This process is repeated until convergence. Each such iteration is a convex optimization problem and we obtain closed form expressions for the updates. Some of the details in this section is skipped for the sake of brevity and relegated to \cite{Kar11c}.

Optimizing \eqref{maxl} with respect to the row-vector $\bo z_k\triangleq \bo Z_{k,\mathcal N_k}\in \mathbb R^{|\mathcal N_k|}$ while keeping all the other elements of $\bo Z$ constant is equivalent \cite{Kar11c} to minimizing
\begin{align}
g(\bo z_k&)=\frac{1}{2}\bo z_k^T\bo \Sigma_k \bo z_k-\log (\bo z_k^T\bo c_k), \label{maxl:pbpo} \\
 &\bo \Sigma_k\in \mathbb R^{|\mathcal N_k|\times |\mathcal N_k|}, \bo c_k\in\mathbb R^{|\mathcal N_k|}, \nonumber
\end{align}
where $\bo \Sigma_k$ denotes the $\mathcal N_k$-clique covariance matrix extracted from $\bo \Sigma$, and the elements of $\bo c_k$ are defined by
\begin{align}
(\bo c_k)_i\triangleq (-1)^{k+j_i}\det(\bo Z_{-k,j_i}), \quad i=1,2,\ldots,|\mathcal N_k|,
\end{align}
with $\bo Z_{-k,j_i}$ denoting the matrix obtained after truncating the $k\sups{th}$ row and $j_i\sups{th}$ column of $\bo Z$. The first-order gradient condition of \eqref{maxl:pbpo} implies $(\bo z_k^T\bo c_k)\bo \Sigma_k\bo z_k=\bo c_k$, solving which one obtains the unique extremum of \eqref{maxl:pbpo},
\begin{align}
\bo z_k^*=\frac{\bo \Sigma_k^{-1}\bo c_k}{\sqrt{\bo c_k^T \bo \Sigma_k^{-1}\bo c_k}}. \label{bcd:update}
\end{align}
That $\bo z_k^*$ is the minimizer follows from the convexity of \eqref{maxl:pbpo} (the Hessian is $(\bo \Sigma_k+(\bo z_k^T\bo c_k)^{-2}\bo c_k \bo c_k^T)$, which is positive definite).

Each rank-one update of the form \eqref{bcd:update} can be efficiently computed using the well-known Woodbury-formula, details of which are relegated to \cite{Kar11c}. Since the overall divergence of \eqref{maxl} decreases at each of the iterations of \eqref{maxl:pbpo}, and the minimum divergence is lower bounded (see equation \eqref{stat:cond}) by
\begin{align}
\sup_{\bo Z\in \mathcal S_A} l(\bo Z)&\ge \sup_{\bo Z\in \mathbb R^{N\times N}} l(\bo Z)= l(\bo L^{-1})= 0,
\end{align}
this iterative algorithm is guaranteed to converge. It may be noted that these kind of iterative techniques are sometimes called block-coordinate-descent or terminal-by-terminal optimization \cite{Gastpar06}.

In the remainder of this paper, we will focus on the application of spatial whitening to distributed estimation.

\section{Example: Bit-Allocation for distributed estimation} \label{sec:bit}
We consider the practical parameter-estimation problem where individual sensors in a network are required to quantize their real-valued local measurements to an appropriate length and send the resulting discrete message to the FC, while the latter combines all the received messages to produce a final estimate \cite{Rib06}. The critical resource that needs to be conserved is the bandwidth or equivalently, the rate of transmission. Assume that the network consisting of $N$ nodes is allowed to transmit only $B$ bits in totality for a one-shot estimation problem. The question then is how to judiciously allocate the $B$ bits among the various sensors such the the resulting distortion of estimate is minimized at the FC \cite{Krasnopeev05}, \cite{Li07}. For the sake of simplicity, we assume that each sensor incurs an equal per-bit cost for transmission.

We would use the quantization and bit allocation framework outlined in \cite{Krasnopeev05}. All observations $x_k$-s are assumed to be bounded to a finite interval $[-U,U]$ and a uniform probabilistic quantization is performed. An observation is quantized with $b_k$-bits as follows. The quantization points ${a_j^{(k)}\in[-U,U],j=1,\ldots,2^{b_k}}$ are uniformly spaced such that $a_{j+1}^{(k)}-a_j^{(k)}=2U/(2^{b_k}-1)\triangleq \Delta_k$. Suppose that $x_k\in[a_j^{(k)},a_{j+1}^{(k)})$. Then $x_k$ is quantized to either $a_{j+1}^{(k)}$ or $a_{j}^{(k)}$ according to
\begin{align}
P(m_k=a_{j}^{(k)})=q, \quad P(m_k=a_{j+1}^{(k)})=1-q, \label{qtzr:unif}
\end{align}
where $m_k$ is the resulting message and $q=(a_{j+1}^{(k)}-x_k)/\Delta_k$.

When the noise is Gaussian and independent across sensors, the subsequent near-optimal strategy \cite{Krasnopeev05} is particularly simple and allocates
\begin{align}
b_k=\text{ROUND}\left[\log_2 \left(1+\frac{1}{\lambda\sigma_k^2} \right)\right] \label{bit:alloc:strat}
\end{align}
bits to the $k\sups{th}$ sensor, where $\sigma_k^2$ is the individual variance,  $\lambda>0$ controls the overall sum of bits $\sum_{k=1}^N b_k=B$ and the rounding is performed to the nearest integer. The idea is that FC broadcasts a lower value of $\lambda$ when a more precise parameter estimate is needed. However, when the noise is correlated, strategy \eqref{bit:alloc:strat} is suboptimal and this is where spatial whitening can be of help. Once we perform a spatial whitening transformation in the observation space, the idea is that we effectively de-correlate the noise without losing any information and hence a strategy like \eqref{bit:alloc:strat} applied on the modified space can still deliver near-optimal performance.

Next we state the distortion metric derived in \cite{Krasnopeev05} which we shall use for comparing the performance of various schemes. For a random variable $\bo y\sim\mathcal{N}(\bo 1 \theta,\bo C)$ that is effectively range limited in $[-U,U]$, the mean-square-error (MSE) for estimating $\widehat \theta$ at FC (when $y_k$ is quantized to $m_k$ using $b_k$ bits) following the scheme in \eqref{qtzr:unif}, is given by
\begin{align}
\text{MSE}(\widehat \theta) \approx \frac{\bo 1^T \bo C^{-1}(\bo C+\bo Q)\bo C^{-1}\bo 1}{(\bo 1^T \bo C^{-1} \bo 1)^2}, \label{mse:mu:hat}
\end{align}
where $\bo Q$ is the diagonal matrix with elements $Q_{kk}=(U^2)/(2^{b_k}-1)^2$. It is assumed that FC is using the optimally weighted fusion rule $\widehat \theta =(\bo 1^T\bo C^{-1} \bo 1)^{-1}\bo 1^T \bo C^{-1}\bo m$ (see \cite{Kay93}) on the quantized observations.

Our simulation setup is as follows. The spatial placement and neighborhood structure is modeled as a Random Geometric Graph $RGG(N,r)$ \cite{Freris10}, where sensors are uniformly distributed over a unit square with communication links present only for pairwise distances of at most $r$. The noise is modeled as an exponentially correlated Gaussian covariance matrix $\bo \Sigma$,
\begin{align}
\bo x \sim \mathcal N(\bo 1 \theta,\bo \Sigma), \quad \bo \Sigma_{i,j}=\sigma_i\sigma_j \alpha^{d_{i,j}},
\end{align}
where $\alpha\in(0,1)$ is indicative of the degree of spatial correlation. A smaller value of $\alpha$ indicates lower correlation with $\alpha\rightarrow 0$ signifying completely independent observations.

\begin{figure}[htb]
\begin{center}
    \includegraphics[width=\figsza \columnwidth]{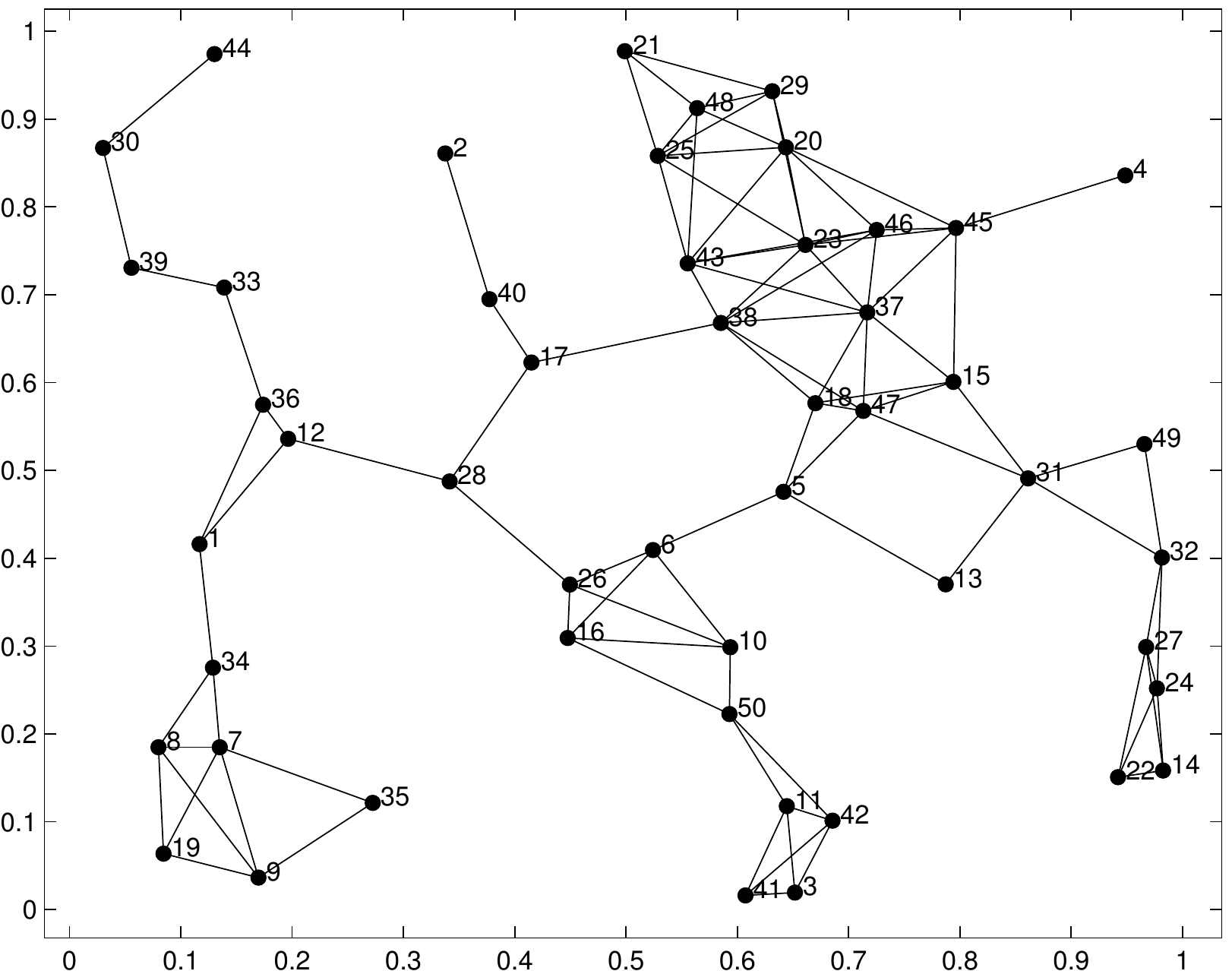}
\vspace{-0.1in}
  \caption{Random Geometric Graph with 50 nodes, used for example in Section \ref{sec:bit}. Edges are shown of pairwise distance less than 0.18.}
\vspace{-0.2in}
  \label{fig:bit:rgg}
  \end{center}
\end{figure}

We consider $N=50$ nodes and the particular RGG used for our simulation is depicted in Figure \ref{fig:bit:rgg}. The individual sensor variances $\sigma_k^2$ are generated by uniform random numbers in the range $[0.5,1.5]$ and the correlation parameter $\alpha=0.02$.
The range-limit of observations is taken as $U=20$.

\begin{figure}[htb]
\begin{center}
    \includegraphics[width=\figszb \columnwidth]{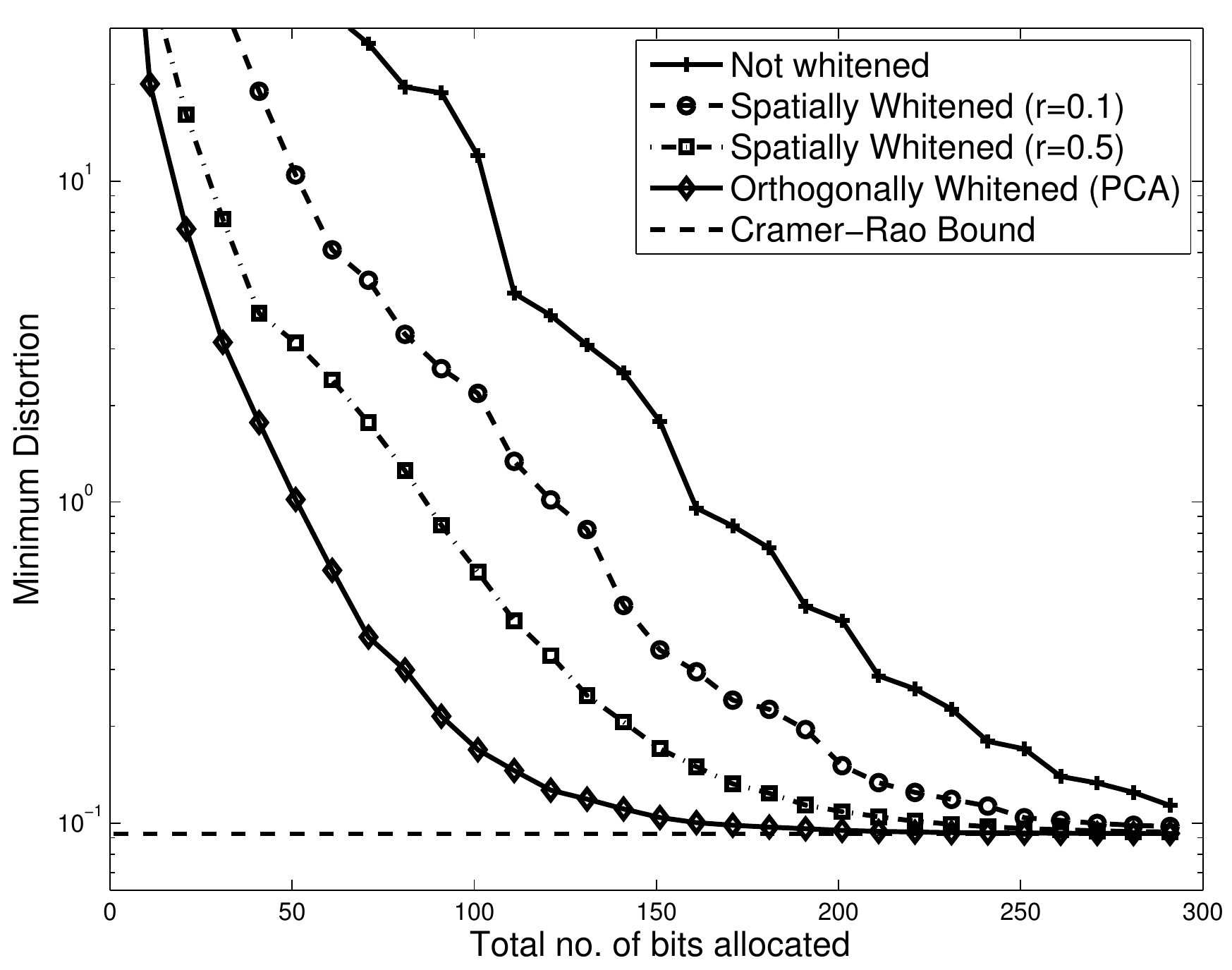}
\vspace{-0.1in}
  \caption{Distortion reduction achieved by spatial whitening.}
\vspace{-0.2in}
  \label{fig:bit}
  \end{center}
\end{figure}

In Figure \ref{fig:bit}, we compare the distortion performance $\text{MSE}(\widehat \theta)$ \eqref{mse:mu:hat} corresponding to the three scenarios when strategy \eqref{bit:alloc:strat} is applied to various transformations of the data. The line labeled \emph{not whitened} corresponds to the naive case of strategy \eqref{bit:alloc:strat} being directly applied to the observation space $\bo x$. Expectedly, the performance of this scheme is suboptimal. In \emph{spatially whitened} cases, we use the transformed variable (see \eqref{def:WD:Z})
\begin{align}
\widetilde{\bo x}=\bo W_r\bo x, \quad \bo W_r=\delta^{-1}(\bo Z_r \bo 1)\bo Z_r, \bo D_r=\delta^{-2}(\bo Z_r \bo 1),
\end{align}
where $\bo Z_r$ is the minimum-divergence solution \eqref{maxl} subject to constraints that $[\bo Z_r]_{ij}=0$ if $d_{ij}>r$. We note that
\begin{align}
\widetilde{\bo x}\sim \mathcal{N}(\bo 1 \theta, \bo W_r \Sigma \bo W_r^T),
\end{align}
which implies that $\widetilde x_k$ possess the same mean as the signal, but corrupted only with \emph{approximately} independent Gaussian noise with variance $\gamma_k^2\triangleq\text{Var}(\widetilde x_k)\approx [\bo D_r]_{k,k}=([\bo Z_r \bo 1]_k)^{-2}$. Strategy \eqref{bit:alloc:strat} is then applied on whitened space $\widetilde{\bo x}$ with $\sigma_k^2$ replaced by $\gamma_k^2$ in Equation \eqref{bit:alloc:strat}. We have shown the performance for $r=0.1$ and $r=0.5$ in Figure \ref{fig:bit}. As the range $r$ increases, we have more whitening and consequently the performance increases. Thirdly, we display the results for \emph{orthogonally whitened (or PCA)} case, where we consider the well known eigenvalue decomposition $\bo \Sigma=\bo U\bo \Lambda \bo U^T$ and consequently the whitening transformation
\begin{align}
\widetilde{\bo x}=\bo D^{-1} \bo U^T\bo x, \quad \bo D\triangleq \delta(\bo U^T \bo 1).
\end{align}
Since PCA fully whitens $\bo \Sigma$ (by definition), its performance is expected to provide a lower bound on that of other schemes. This is confirmed by Figure \ref{fig:bit}. However, since the weights in PCA are not designed to be zero for sensors that are far apart, such a transformation may be impossible to realize in a power constrained network and hence not realistic. Finally, the Cramer-Rao lower bound $\text{CRB}=1/(\bo 1^T \bo \Sigma^{-1} \bo 1)$ is also displayed, which confirms that in the asymptotic regime with sufficient quantization bits per sensor, all these schemes perform identically.

\section{Conclusion}
In this paper, we have considered a two-stage framework for distributed signal processing in the presence of spatially correlated data. The first stage is designed to whiten the observation space by communicating only with neighboring sensors. In the second stage, each sensor encodes these whitened observations following well-known strategies derived using conditional independence assumption. We consider the example of bit-allocation for distributed estimation to demonstrate the potential applicability of this framework. Many research questions remain to be addressed. Some of them are efficient computation of the spatial whitening transformation, cost considerations for the whitening stage, extension of the framework to vector parameter scenarios and potential applicability in hypothesis testing problems.


\bibliographystyle{IEEEtran}
\bibliography{thesis}

\end{document}